\documentclass{bioinfo}
\copyrightyear{2015}
\pubyear{2015}

\usepackage{amsmath}
\usepackage[ruled,vlined]{algorithm2e}

\SetCommentSty{mycommfont}
\SetKwComment{Comment}{$\triangleright$\ }{}
\usepackage{natbib}

\begin{document}
\firstpage{1}

\title[Error Correction for Illumina Data]{Correcting Illumina sequencing errors for human data}

\author[Li]{Heng Li}

\address{Broad Institute, 75 Ames Street, Cambridge, MA 02142, USA}

\history{Received on XXXXX; revised on XXXXX; accepted on XXXXX}
\editor{Associate Editor: XXXXXXX}
\maketitle

\begin{abstract}
\section{Summary:} We present a new tool to correct sequencing errors in
Illumina data produced from high-coverage whole-genome shotgun resequencing. It
uses a non-greedy algorithm and shows comparable performance and higher
accuracy in an evaluation on real human data. This evaluation has the most
complete collection of high-performance error correctors so far.

\section{Availability and implementation:} https://github.com/lh3/bfc

\section{Contact:} hengli@broadinstitute.org
\end{abstract}

\section{Introduction}

Error correction is a process to fix sequencing errors on a sequence read
by using other overlapping reads that do not contain the errors. Many \emph{de
novo} assemblers, in particular short-read assemblers for large genomes, use
error correction to reduce the complexity of the assembly graph such that the
graph can be fitted to limited RAM. Error correction was first expressed
as the \emph{spectrum alignment problem}~\citep{Pevzner:2001vn}, whereby we take
a set of trusted $k$-mers and attempt to find a sequence with minimal
corrections such that each $k$-mer on the corrected sequence is trusted.
The majority of error correctors are based on this idea and take a greedy
approach to solving this problem.  They make a correction based on the local
sequence context and do not revert the decision. They may not find the sequence
with the minimal corrections. We worried that the greedy strategy might affect
the accuracy given reads from a repeat-rich diploid genome, so derived a new
algorithm. It is optimal provided that we know there is an error-free $k$-mer
in the read.

\begin{methods}
\section{Methods}
Algorithm~1 is the key component of BFC. It defines a \emph{state}
of correction as a 4-tuple $(i,W,\mathcal{C},p)$, which consists of the
position $i$ of the preceding base, the last \mbox{($k$-1)--mer} $W$ ending at
$i$, the set $\mathcal{C}$ of previous corrected positions and bases (called a
\emph{solution}) up to $i$, and the penalty $p$ of solution $\mathcal{C}$. BFC
keeps all possible states in a priority queue $\mathcal{Q}$. At each iteration,
it retrieves the state $(i,W,\mathcal{C},p)$ with the lowest penalty $p$ (line~1) and
adds a new state $(i+1,W[1,k-2]\circ a,\mathcal{C}',p')$ if $a$ is the read
base or $W\circ a$ is a trusted $k$-mer. If the first $k$-mer in $S$ is
error free and we disallow untrusted $k$-mers by removing line~3, this
algorithm finds the optimal solution to the spectrum alignment problem.

\begin{algorithm}[ht]
\DontPrintSemicolon
\footnotesize
\KwIn{K-mer size $k$, set $\mathcal{H}$ of trusted $k$-mers, and one string $S$}
\KwOut{Set of corrected positions and bases changed to}
\BlankLine
\textbf{Function} {\sc CorrectErrors}$(k, \mathcal{H}, S)$
\Begin {
	$\mathcal{Q}\gets${\sc HeapInit}$()$\Comment*[r]{$\mathcal{Q}$ is a priority queue}
	{\sc HeapPush}$(\mathcal{Q}, (k-2, S[0,k-2], \emptyset, 0))$\Comment*[r]{0-based strings}
	\While{$\mathcal{Q}$ is not empty} {
		\nl$(i, W, \mathcal{C}, p)\gets${\sc HeapPopBest}$(\mathcal{Q})$\Comment*[r]{current best state}
		$i\gets i+1$\;
		\lIf{$i=|S|$} { {\bf return} $\mathcal{C}$ \Comment*[f]{reaching the end of $S$}}
		\nl$\mathcal{N}\gets\{(i,{\rm A}),(i,{\rm C}),(i,{\rm G}),(i,{\rm T})\}$\Comment*[r]{set of next bases}
		\ForEach (\Comment*[f]{try all possible next bases}) {$(j,a)\in\mathcal{N}$} {
			$W'\gets W\circ a$\Comment*[r]{``$\circ$'' concatenates strings}
			\uIf (\Comment*[f]{no correction}) {$i=j$ {\bf and} $a=S[j]$} {
				\eIf (\Comment*[f]{good read base; no penalty}) {$W'\in\mathcal{H}$} {
					{\sc HeapPush}$(\mathcal{Q}, (j,W'[1,k-1],\mathcal{C}, p))$\;
				} (\Comment*[f]{bad read base; penalize}) {
					\nl{\sc HeapPush}$(\mathcal{Q}, (j,W'[1,k-1],\mathcal{C}, p+1))$\;
				}
			} \ElseIf (\Comment*[f]{make a correction with penalty}) {$W'\in \mathcal{H}$} {
				\nl{\sc HeapPush}$(\mathcal{Q}, (j,W'[1,k-1],\mathcal{C}\cup\{(j,a)\},p+1))$\;
			}
		}
	}
}
\caption{Error correction for one string in one direction}
\end{algorithm}

It is possible to modify the algorithm to correct insertion and deletion
errors (INDELs) by augmenting the set of the ``next bases'' at line~2 to:
$$\mathcal{N}=\{(j,a)|j\in\{i-1,i\},a\in\{{\rm A},{\rm C},{\rm G},{\rm T}\}\}\cup\{(i,\epsilon)\}$$
In this set, $(i,a)$ substitutes a base at position $i$, $(i,\epsilon)$ deletes
the base and $(i-1,a)$ inserts a base $a$ before $i$. We have not
implemented this INDEL-aware algorithm because such
errors are rare in Illumina data.

The worse-case time complexity of Algorithm~1 is exponential in the length of
the read. In implementation, we use a heuristic to reduce the search space by
skipping line~4 if the base quality is 20 or higher (Q20) and the $k$-mer
ending at it is trusted, or if five bases or two Q20 bases have been corrected
in the last 10bp window. If BFC still takes too many iterations before finding
an optimal solution, it stops the search and marks the read uncorrectable.

Given a read, BFC finds the longest substring on which each $k$-mer is trusted. It
then extends the substring to both ends of the read with Algorithm~1. If a read
does not contain any trusted $k$-mers, BFC exhaustively enumerates all $k$-mers
one-mismatch away from the first $k$-mer on the read to find a trusted
$k$-mer. It marks the read uncorrectable if none or multiple trusted $k$-mers
are found this way.

We provided two related implementations of Algorithm~1, BFC-bf and BFC-ht.
BFC-bf uses KMC2~\citep{kmc2} to get exact $k$-mers counts and then keeps
$k$-mers occurring three times or more in a blocked bloom
filter~\citep{DBLP:conf/wea/PutzeSS07}. BFC-ht uses a combination of bloom
filter and in-memory hash table to derive approximate $k$-mer
counts~\citep{Melsted:2011bh} and counts of $k$-mers consisting of Q20 bases.
We modified Algorithm~1 such that missing trusted high-quality $k$-mers incurs
an extra penalty. This supposedly helps to correct systematic sequencing errors
which are recurrent but have lower base quality.

%To take advantage of multiple CPU cores, we implemented a blocked bloom
%filter~\citep{DBLP:conf/wea/PutzeSS07} with a 1-byte spin lock for every
%63-byte block to prevent concurrent modifications to the same block.  For human
%data, we used an array of 16 million (=$2^{24}$) hash tables to keep $k$-mer
%counts. Each hash table has a spin lock. With a good hash function, locking is
%infrequent. We also put computing and I/O on different threads.  The strategy
%is particularly useful when I/O is slow.

\end{methods}

\section{Results and Discussions}

We evaluated BFC along with BBMap-34.38 (\mbox{http://bit.ly/bbMap}),
BLESS-v0p23~\citep{Heo:2014aa}, Bloocoo-1.0.4~\citep{Drezen:2014aa},
fermi2-r175~\citep{Li:2012fk}, Lighter-20150123~\citep{Song:2014aa},
Musket-1.1~\citep{Liu:2013ac} and SGA-0.9.13~\citep{Simpson:2012aa} on real
data (Table~1). We ran the tools on a Linux server with 20 cores of Intel
E5-2660 CPUs and 128GB RAM. Precompiled binaries are available through
http://bit.ly/biobin and the command lines were included in the BFC source code
package (http://bit.ly/bfc-eval).  Notably, BLESS only works with uncompressed
files. The rest of tools were provided with gzip'd files as input. We have also
tried AllPaths-LG~\citep{Gnerre:2011ys}, Fiona~\citep{Schulz:2014aa} and
Trowel~\citep{Lim:2014aa}, but they require more RAM than our machine.
QuorUM-1.0.0~\citep{Zimin:2013aa} always trims reads, making it hard to be
compared to others which keep full-length reads.

\begin{table}[t]
\processtable{Performance of error correction}
{\footnotesize
\begin{tabular}{lcrrccrr}
\toprule
Prog.     & $k$ & Time  & RAM   & Perfect&Chim.& Better & Worse \\
\midrule
raw data  & --  & --    & --    & 2.40M  & 12.4k  & --     & -- \\
BBMap     & 31  &{\bf 3h22m}&33.0G& 2.78M& 12.4k  & 505k   & 19.2k \\
%BFC-ht    & 37  & 6h33m & 63.5G & 3.04M  & 12.5k  & 830k   & 9.7k \\
BFC-ht    & 31  & 7h15m & 83.5G & 3.03M  &13.6k   & 816k   & 10.8k \\
%BFC-ht    & 33  & 6h15m & 61.4G & 3.03M  &13.2k& 822k   & 10.1k \\
BFC-ht    & 55  & 5h51m & 67.9G &{\bf 3.05M}&11.7k& 830k   &{\bf 9.0k}\\
%BFC-ht    &55/33& 9h18m & 67.9G &{\bf 3.07M}&11.8k&{\bf 861k}&9.5k\\
BFC-bf    & 31  & 7h32m & 23.3G & 3.01M  & 13.1k  & 783k   & 9.2k \\
BFC-bf    & 55  & 4h41m & 23.3G &{\bf 3.05M}&11.8k& 819k   & 11.4k \\
BLESS     & 31  & 6h31m & 22.3G & 2.91M  & 13.1k  & 674k   & 20.8k \\
BLESS     & 55  & 5h09m & 22.3G & 3.01M  &{\bf 11.5k}& 775k& 10.3k \\
Bloocoo   & 31  & 5h52m &{\bf 4.0G}&2.88M& 14.1k  & 764k   & 31.5k  \\
Fermi2    & 29  &17h14m & 64.7G & 3.00M  & 17.7k  &{\bf 849k}&42.8k \\
Lighter   & 31  & 5h12m & 13.4G & 2.98M  & 13.0k  & 756k   & 30.1k  \\
Musket    & 27  &21h33m & 77.5G & 2.94M  & 22.5k  & 790k   & 36.3k  \\
SGA       & 55  &48h40m & 35.6G & 3.01M  & 12.1k  & 755k   & 12.8k  \\
\botrule
\end{tabular}}{4.45 million pairs of $\sim$150bp reads were downloaded from
BaseSpace, under the sample ``NA12878-L7'' of project ``HiSeq X Ten: TruSeq
Nano (4 replicates of NA12878)'', and were corrected together. On a subset of
two million randomly sampled read pairs, the original and the corrected
sequences were mapped to hs37d5 (http://bit.ly/GRCh37d5) with
BWA-MEM~\citep{Li:2013aa}.  A read is said to become \emph{better} (or
\emph{worse}) if the best alignment of the corrected sequence has more (or
fewer) identical bases to the reference genome than the best alignment of the
original sequence. The table gives $k$-mer size (maximal size used for Bloocoo,
fermi2, Lighter and Musket), the wall-clock \emph{time} when 16 threads are
specified if possible, the peak \emph{RAM} measured by GNU time, number of
corrected reads mapped \emph{perfectly}, number of \emph{chimeric} reads,
number of corrected reads becoming \emph{better} and the number of reads
becoming \emph{worse} than the original reads. For each metric, the best tool
is highlighted in the bold fontface.}
\end{table}

As is shown in the table, BBMap is the fastest. BFC, BLESS, Bloocoo and Lighter
are comparable in speed. Bloocoo is the most lightweight. Other bloom filter
based tools, BFC-bf, BLESS and Lighter, also have a small memory footprint.
Most evaluated tools have broadly similar accuracy.  BFC-ht is more accurate
than BFC-bf overall, suggesting retaining high-quality $k$-mers helps error
correction; both BFC implementations are marginally better in this evaluaton,
correcting more reads with fewer or comparable overcorrections when a similar
$k$-mer length is in use, which potentially demonstrates that a non-greedy
algorithm might work better, though subtle differences in heuristics and hidden
thresholds between the tools could also play a role.  We should note that it is
possible to tune the balance between accuracy, speed and memory for each tool.
We have not fully explored all the options.

In the table, error correctors appear to be faster and more accurate when
longer $k$-mers are in use. A possible explanation is that longer $k$-mers
resolve more repeat sequences and also reduce the search space. However, when
we use BFC-ht to correct errors in this dataset, fermi2~\citep{Li:2012fk}
derived longer contigs and better variant calls with shorter $k$-mers. We
speculate that this observation is caused by reduced $k$-mer coverage firstly
because there are fewer long $k$-mers on each read and secondly because longer
$k$-mers are more likely to harbor errors. The reduced $k$-mer coverage makes
it harder to correct errors in regions with low coverage and thus increases the
chance of breaking contigs.  To take advantage of both shorter and longer
$k$-mers, we have also tried a two-round correction strategy with two $k$-mer
sizes. The strategy leads to better numbers in the table (861k reads corrected
to be better and 9.5k worse) at the cost of speed, but does not greatly improve
the assembly. We will focus on understanding the interaction between error
correctors and assemblers in future works.

\section*{Acknowledgement}
\paragraph{Funding\textcolon} NHGRI U54HG003037; NIH GM100233

\bibliography{bfc}
\end{document}